\documentclass[12pt]{article}
\usepackage{amsmath}
\usepackage{amssymb}
\usepackage{graphicx}
\usepackage{bm}
\usepackage{upgreek}

\begin{document}

{\noindent \bf  VISCOSITY-INDUCED CROSSING OF THE PHANTOM DIVIDE IN THE DARK COSMIC FLUID}
\vspace{1cm}

\begin{center}
Iver Brevik\footnote{iver.h.brevik@ntnu.no}

\bigskip
Department of Energy and Process Engineering,\\ Norwegian University of Science and Technology, Trondheim, Norway

\end{center}

\begin{abstract}

Choosing various natural  forms for the equation-of-state parameter $w$ and the bulk viscosity $\zeta$, we discuss how it is possible for a dark energy fluid to slide from the quintessence region across the divide $w=-1$ into the phantom region, and thus into a Big Rip future singularity. Different analytic forms for $\zeta$, as powers of the scalar expansion, are suggested and compared with experiments.

\end{abstract}

\begin{center}

\end{center}

\section{Introduction}

The discovery of the accelerated universe \cite{riess98,perlmutter99} has led to new concepts and ideas in cosmology, in particular, the concept of a dark energy (for a recent review, see \cite{bamba12}). A characteristic feature of dark energy is the pressure of a negative pressure (i.e., a positive tensile stress) in the cosmic fluid. About 73\% of the total mass/energy in the universe consists of dark energy, whereas only 27\% consists of a combination of dark matter and baryonic matter. Even modifications of gravity theory itself is a topic that has attracted considerable interest (a review of this kind of theory can be found in  \cite{nojiri11}).

Usually, one takes the equation of state for the cosmic fluid in the homogeneous form
\begin{equation}
p=w\rho, \label{1}
\end{equation}
with $p$ the pressure and $\rho$ the mass/energy. Thus $w=0$ corresponds to the pressure-less fluid. When $w<0$, strange thermodynamic effects are encountered. Thus  when the borderline $w=-1/3$ is crossed, marking the transition into the so-called quintessence region, the strong energy condition $\rho+3p \geq 0$ becomes violated \cite{hawking73}. Even more peculiar properties are encountered when the borderline $w=-1$ (called the phantom divide) is crossed. A characteristic feature here is that a singularity of the universe may occur, in a finite span of time. It is called a Big Rip. The possibility for such a fate of our universe  was first pointed out by Caldwell {\it et al.} about 10 years ago \cite{caldwell02,caldwell03}, and has later been studied by a number of authors; cf., for instance,  \cite{nojiri04}.  There are actually several variants of future singularity theories, implying a more 'soft' approach of the universe to this particular limit.  Thus scenarios of Little Rip \cite{frampton11}, Pseudo Rip \cite{frampton12}, and Quasi Rip \cite{wei12} have recently been discussed in the literature. Our references to the literature are here very limited; a much more extensive overview is given, for instance,  in the recent paper \cite{brevik13}.

 Recent astronomical observations indicate that the value of $w$ lies close to $-1$,
\begin{equation}
w=-1.04^{+0.09}_{-0.10}; \label{2}
\end{equation}
cf. \cite{nakamura10}, and so a detailed analysis of the behavior of a dark energy fluid is of obvious physical interest. In view of the dominance of the dark energy fluid component in the universe we shall for simplicity consider a model containing one dark component only. Moreover, as an essential point we shall analyze the influence from a {\it viscosity} in the cosmic fluid. Most of the earlier cosmological theories  have assumed the fluid to be nonviscous. From a hydrodynamicist's point of view this is actually somewhat surprising, since viscosity effects so often play a role in ordinary fluid mechanics. In accordance with common usage we shall take the universe to be spatially isotropic, meaning that it is the bulk viscosity, called $\zeta$, and not the shear viscosity, that becomes relevant. One important property on which we shall focus attention in the following, is that when $\zeta$ is taken to be positive it becomes possible for the fluid to slide from the quintessence region (i.e. $-1<w<-1/3$) through the phantom divide into the phantom region and thereafter into the future singularity. This was first pointed out in \cite{brevik05}. Obviously, the magnitude of $\zeta$ will be important for this transition process. Specifically, we shall consider the following two points:
\begin{itemize}
\item  What is the influence from the equation-of-state parameter $w$ for this process?
\item What is the influence from the use of different  analytic forms for the bulk viscosity $\zeta$?
\end{itemize}
These questions will be dealt with  in sections 2 and 3 below. Some comparison with astronomical observations will also be made in section 4.

  Some more references to papers on viscous cosmology are  \cite{padmanabhan87} and \cite{gron90} (these are early papers, the latter being an extensive review up to 1990). Later works can be found in \cite{brevik94,zimdahl96,mostafapoor11,dou11,brevik/gron13}. We also mention two works where transition through the phantom barrier was considered in a more general context \cite{nojiri05,nojiri06}.

\section{General formalism, and the case when $w$ is constant}

 Let $g_{\mu\nu}$ be the general metric such that the  diagonal components are $(-,+,+,+)$ in the Minkowski case, and let $h_{\mu\nu}=g_{\mu\nu}+U_{\mu}U_{\nu}$ be the projection tensor with $U_\mu$ the fluid's four-velocity. Then, since the shear viscosity is assumed to be zero, the energy-momentum tensor can be written in the simple form
\begin{equation}
T_{\mu\nu}=\rho U_\mu U_\nu +(p-\zeta \theta)h_{\mu\nu}, \label{3}
\end{equation}
where $\theta={U^\lambda}_{;\lambda}$ is the scalar expansion.

The FRW metric in comoving coordinates is
\begin{equation}
ds^2=-dt^2+a^2(t)d{\bf x}^2, \label{4}
\end{equation}
where $a(t)$ is the scale factor. In this metric $\theta=3H$, where $H=\dot{a}/a$ is the Hubble parameter. Defining $\kappa=8\pi G$ we can write the Friedmann equations as
\begin{equation}
\theta^2=3\kappa \rho, \label{5}
\end{equation}
\begin{equation}
2\dot{\theta}+\theta^2=-3\kappa(p-\zeta \theta). \label{6}
\end{equation}
The energy conservation equation ${T^{0\nu}}_{;\nu}=0$ implies
\begin{equation}
\dot{\rho}+(\rho+p)\theta=\zeta \theta^2. \label{7}
\end{equation}

Let us now consider the equation of state for the dark energy fluid, and first assume that
\begin{equation}
w=-1-\alpha, \label{8}
\end{equation}
where $\alpha$ is a constant. Thus $\alpha=0$ corresponds to the presence of a cosmological constant $\Lambda$ in conventional relativity, while $\alpha>0$ corresponds to the phantom region.

We can now derive the governing equations for the scalar expansion, or equivalently, for the energy density. The governing equation for $\rho$, taking (\ref{8}) into account, becomes
\begin{equation}
\dot{\rho}-\sqrt{3\kappa}\,\alpha  \rho^{3/2}-3\kappa\rho \zeta(\rho)=0, \label{9}
\end{equation}
which has the solution
\begin{equation}
t=\frac{1}{\sqrt{ 3 \kappa}} \frac{1}{\alpha}\int_{\rho_0}^\rho \frac{d\rho}{\rho^{3/2}[1+\sqrt{3\kappa/\rho} \,\zeta(\rho)/\alpha]}. \label{10}
\end{equation}
Here the integration is taken from present time $t=0$ when the density is $\rho_0$,   to an arbitrary time $t$ in the future.

We consider now different assumptions for the form of the bulk viscosity.

{\it (i)  $\zeta$ equal to a constant.} Let us assume
\begin{equation}
\zeta =\zeta_0, \label{11}
\end{equation}
with $\zeta_0$ a constant. From the above equations we get
\begin{equation}
\theta(t)=\theta_0\frac{e^{t/t_c}}{1-\frac{1}{2}\alpha \theta_0t_c(e^{t/t_c}-1)}, \label{12}
\end{equation}
where $\theta_0$  is the present-time expansion and $t_c$ the 'viscosity time'
\begin{equation}
t_c=\frac{2}{3\kappa \zeta_0}. \label{13}
\end{equation}
The density will vary with time as
\begin{equation}
\rho(t)=\rho_0\frac{e^{2t/t_c}}{\left[ 1-\frac{1}{2}\alpha \theta_0t_c(e^{t/t_c}-1)\right]^2}. \label{14}
\end{equation}
We can now make the following important observation: If the universe starts from a state lying in the phantom region, $\alpha >0$, it will inevitably be developing into a future singularity of the Big Rip type, at a finite time
\begin{equation}
t_s=t_c\ln \left( 1+\frac{2}{\alpha \theta_0t_c}\right). \label{15}
\end{equation}
By contrast, if it starts from the quintessence region, $\alpha <0$, the universe  will never encounter a future singularity. Both $\theta (t)$ and $\rho(t)$ tend to finite values as $t\rightarrow \infty$. The scale factor $a(t)\rightarrow 0$.

{\it (ii) $\zeta$ proportional to $\theta$.} Let us now make the ansatz
\begin{equation}
\zeta(\rho)=\tau_1 \theta=\tau_1\sqrt{3\kappa \rho}. \label{16}
\end{equation}
This is  physically reasonable, as the viscosity may be expected to increase during the violent motions of the cosmic fluid towards the future singularity. Equation (\ref{10}) yields now
\begin{equation}
t=\frac{1}{\sqrt{3\kappa}}\frac{2}{\alpha+3\kappa \tau_1}\left( \frac{1}{\sqrt{\rho_0}}-\frac{1}{\sqrt \rho}\right). \label{17}
\end{equation}
From this we see the following: If the universe starts from the phantom region $\alpha >0$ at $t=0$, it will inevitably end up in a future Big Rip singularity at a finite time, irrespective of the value of the parameter $\tau_1$. If the the starting point lies in the quintessence region, however, the fate of the universe will depend on how viscous the universe is. The point is whether $-|\alpha|+3\kappa \tau_1$ is positive or negative. Thus, if $\tau_1$ is larger than a critical value given by
\begin{equation}
\tau_{1,\rm crit}=\frac{|\alpha|}{3\kappa}, \label{18}
\end{equation}
the Big Rip singularity $(\rho =\infty)$ occurs. If this condition is not met,  $\rho(t) \rightarrow 0$ as $t\rightarrow \infty$.

This property of the universe was pointed out also earlier, in \cite{brevik05}.

{\it (iii) $\zeta$ proportional to $\theta^2$.} A natural generalization of the above ansatz is to consider the case when $\zeta(\rho)$ is proportional to the square of the expansion,
\begin{equation}
\zeta(\rho)=\tau_2\theta^2=3\kappa \rho \tau_2. \label{19}
\end{equation}
We assume that $\tau_2$, like $\tau_1$ above, are positive quantities,  as  viscosities should be positive for  dissipative processes. (In practice, one would expect that a linear combination of (\ref{16}) and (\ref{19}) occurs, but for simplicity we consider here the ansatz (\ref{19}) alone.) From (\ref{10}) we now get
\[
t=\frac{2}{\sqrt{3\kappa}}\int_{\sqrt {\rho_0}}^{\sqrt \rho}\frac{dx}{x^2(\alpha +A x)} \]
\begin{equation}
=\frac{2}{\sqrt{3\kappa}}\left\{ \frac{1}{\alpha}\left(\frac{1}{\sqrt {\rho_0}}-\frac{1}{\sqrt \rho}\right)+
\frac{A}{\alpha^2}\ln \left(\frac{\sqrt{\rho_0}}{\sqrt\rho}\frac{\alpha +A\sqrt{\rho}}{\alpha +A\sqrt{\rho_0}}\right)\right\}. \label{20}
\end{equation}
with
\begin{equation}
A= (3\kappa)^{3/2}\tau_2. \label{21}
\end{equation}
 If $\alpha >0$ (phantom region) the universe thus runs into a Big Rip singularity, $\rho=\infty$, at a finite time
 \begin{equation}
 t_{\rm rip}=\frac{2}{\sqrt{3\kappa}}\frac{1}{\alpha}\frac{1}{\sqrt{\rho_0}}; \label{22}
 \end{equation}
 the logarithmic term in (\ref{20}) fades away.

 If $\alpha <0$ (quintessence region), the situation becomes however complicated. If $-|\alpha|+A\sqrt{\rho_0}>0$  at the initial instant $t=0$, the logarithmic term in (\ref{20}) fades away when $\rho \rightarrow \infty$, but the expression for $t$ becomes negative because of the factor $1/\alpha$ in the first term in (\ref{20}). That is  unacceptable, since we are looking at the development of the universe in the future only.    If $-|\alpha|+A\sqrt{\rho_0}<0$, a logarithmic singularity $(t\rightarrow -\infty)$ is encountered when  $\sqrt{\rho}=|\alpha|/A$. We conclude that the case $\alpha<0$ is  hardly of physical interest here.

 \section{A more general form for the equation-of-state parameter $w=w(\rho)$}

 We now make some remarks on   the case when $w$ is still taken to be a function of $\rho$, but has a more general form. Let us start with the ansatz
 \begin{equation}
 p=-\rho-\alpha {\rho}^\beta, \label{23}
 \end{equation}
 where $\alpha$ and $\beta$ are unspecified  constants to begin with. Here $\beta$ is nondimensional, while the dimension of $\alpha$ is $[\alpha]={\rm cm}^{4(\beta-1)}$ in geometric units. The expression (\ref{23}) means that
 \begin{equation}
 w=-1-\alpha { \rho}^{\beta-1}. \label{24}
 \end{equation}
 The previous case (\ref{8}) corresponds to the choice $\beta=1$.

Equation (\ref{10}) becomes now replaced by
\begin{equation}
t=\frac{1}{\sqrt{3\kappa}}\int_{\rho_0}^\rho  \frac{d\rho}{{\rho}^{\beta +1/2} \left[\alpha+\sqrt{3\kappa}\,\zeta(\rho){\rho}^{-\beta+1/2}\right]}. \label{25}
\end{equation}
Looking for a mathematically simple and at the same time a physical reasonable form for the viscosity, we see that the following form
\begin{equation}
\zeta(\rho)=\tau \theta^{2\beta-1}=\tau(3\kappa \rho)^{\beta-1/2}, \label{26}
\end{equation}
with $\tau$ a positive constant, is most natural. Then for $\beta=1$ the case {\it (ii)} in the previous section is recovered with $\tau=\tau_1$, and for $\beta=3/2$ the case {\it (iii)} is recovered with $\tau=\tau_2$. We see that with (\ref{26}) the expression between square parentheses in (\ref{25}) becomes independent of $\rho$, and so
\[ t=\frac{1}{\sqrt{3\kappa}}\,\frac{1}{\alpha+(3\kappa)^\beta \tau}
\int_{\rho_0}^\rho \frac{d\rho}{\rho^{\beta+1/2}} \]
\begin{equation}
=\frac{1}{\sqrt{3\kappa }}\,\frac{2}{2\beta-1}\,\frac{1}{\alpha+(3\kappa)^\beta \tau}
\left(\frac{1}{\rho_0^{\beta-1/2}}- \frac{1}{\rho^{\beta-1/2}} \right).
 \label{27}
\end{equation}
In order to obtain a convergent integral over $\rho$ when the  upper limit is chosen as $\rho=\infty$, one must have $\beta>1/2$. For a Big Rip to occur in a finite time $t$ one must in addition have the condition $\alpha +(3\kappa)^\beta \tau >0$ satisfied.  We see that the universe possesses the same ability to slide through the phantom divide $w=-1$ as we  saw before: The universe may  start from a point in the quintessence region, $\alpha<0$, and yet run into a Big Rip singularity if the coefficient $\tau$ in (\ref{26}) is large enough. The condition for Big Rip is seen from (\ref{27}) to be
\begin{equation}
\tau_{\rm crit}>\frac{|\alpha|}{(3\kappa)^\beta}. \label{28}
\end{equation}

\section{Conclusions}

Choosing various forms for the equation-of-state parameter $w=w(\rho)$ and the bulk viscosity $\zeta=\zeta(\rho)$, our main objective has been to discuss the possibilities the dark energy universe has to slide from the quintessence region $w>-1$ into the phantom region $w<-1$ and thus into the future Big Rip singularity. The sliding process is thus viscosity-generated.

If $w$ is assumed constant, set equal to $-1-\alpha$ in (\ref{8}), the universe possesses this property in a natural way if $\zeta(\rho)$ is taken to be proportional to the scalar expansion $\theta$; cf. (\ref{16}).

If $w(\rho)$ has the more complicated form (\ref{24}) the same property persists, if $\zeta(\rho)$ is taken to have the form (\ref{26}), what is a natural generalization. For a Big Rip to occur, the coefficient $\beta$ in (\ref{24}) must be larger than 1/2.

Finally, it is of physical interest to investigate possible relationships between the assumptions made above, and observations in cosmology.  As we know, for an imperfect fluid the viscosity is generated by molecular interactions and can be represented as a functions of macroscopic thermodynamical variables such as temperature. Thus, we can assume the form  $\zeta =\zeta(T)$. It is natural to make use of conventional kinetic theory. One possibility is to adopt  the Chapman-Enskog formula for a dilute fluid (cf. the Chapman-Cowling volume \cite{chapman95}) according to which, for low temperatures ($T<300$ K),  we can approximate $\zeta \propto T^{1/2}$. Another possibility is to choose the Sutherland formula, implying $\zeta \propto T^{3/2}$. An analysis if this sort was recently given by  Wang and Meng \cite{wang13}, comparing with astronomical observations. The temperature was identified with that of cosmic microwave radiation (CMB),
\begin{equation}
T(z)=T_0(1+z), \label{29}
\end{equation}
with $T_0=2.73$ K the present CMB temperature and $z$ the redshift. One could thus write
\begin{equation}
\zeta=\zeta_0[T_0(1+z)]^\alpha, \label{30}
\end{equation}
with $\zeta_0$ an effective constant and $\alpha=1/2$ or $\alpha=3/2$ in the Chapman-Enskog or Sutherland cases, respectively.

Comparing with different observational data sets, Wang and Meng were able to give approximate values for the quantity $ 12\pi G\zeta_0T_0^\alpha$ for the two cases mentioned. We reproduce here the values inferred from the SNe Ia data:
\begin{equation}
12\pi G\zeta_0T_0^\alpha =\left\{ \begin{array}{ll}
0.87 & \alpha=1/2 \\
1.61 & \alpha=3/2
\end{array}
\right. \label{31}
\end{equation}
These numerical estimates are of obvious  physical interest. For our purpose here the main issue is  however to make a comparison between the exponents: Assume first that the universe is flat and  matter dominated, $\rho \propto a^{-3}$, so that
\begin{equation}
a(t)=2.3\times 10^{-12}t^{2/3}, \quad T(t)=10^{12}t^{-2/3}~\rm K. \label{32}
\end{equation}
Then,
\begin{equation}
T\propto 1/a  \propto t^{-2/3}, \quad \theta=3H= 2/t. \label{33}
\end{equation}
This means that  $T\propto \theta^{2/3}$, so that the ansatz (\ref{30}) above implies
\begin{equation}
\zeta \propto \theta^{2\alpha/3}. \label{34}
\end{equation}
We thus see that our first option (\ref{16}) above, $\zeta \propto \theta$, agrees with (\ref{34}) when $\alpha=3/2$. This is actually the Sutherland case. This correspondence is physically satisfactory, since classical kinetic theory obviously deals with  systems composed of matter particles.  Our second option in  Eq.~(\ref{19}),  $\zeta \propto \theta^2$, corresponds to $\alpha=3$ and is probably of less physical interest.

If on the other hand the universe is taken to be flat and radiation dominated, $\rho \propto a^{-4}$,   we have
\begin{equation}
a(t)=2.2\times 10^{-10}t^{1/2}, \quad T(t)=10^{10}t^{-1/2}~\rm K, \label{35}
\end{equation}
so that
\begin{equation}
T\propto 1/a \propto t^{-1/2}, \quad \theta=3H=3/(2t). \label{36}
\end{equation}
Then $T\propto \theta^{1/2}$, so that according to (\ref{30})
\begin{equation}
\zeta \propto \theta^{\alpha/2}. \label{37}
\end{equation}
In order to get $\zeta \propto \theta$, our preferred option above, we thus have to set $\alpha =2$, not so very far from the Sutherland value 3/2 after all.  The case $\zeta \propto \theta^2$ corresponds now to $\alpha=4$.

\bigskip
{\bf Acknowledgment}

\bigskip
I thank Sergei D. Odintsov for valuable information.


\end{document}